\documentstyle[12pt]{article}
\def\bbox{{\,\lower0.9pt\vbox{\hrule \hbox{\vrule height 0.2 cm

\hskip 0.2 cm

\vrule  height 0.2 cm}\hrule}\,}}

\def\bbox{{\,\lower0.9pt\vbox{\hrule \hbox{\vrule height 0.2 cm

\hskip 0.2 cm

\vrule  height 0.2 cm}\hrule}\,}}
\begin{document}
\setlength{\unitlength}{1mm}
\title{{\hfill {\small  } } \vspace*{2cm} \\
Black Hole Thermodynamics, Induced Gravity and \\
Gravity in Brane Worlds}
\author{\\
D.V. Fursaev\thanks{e-mail: fursaev@thsun1.jinr.ru}
${}^{1}$
\date{}}
\maketitle
\noindent  
 \\ 
$^{1}${\em Joint Institute for Nuclear Research, Bogoliubov
Laboratory of Theoretical Physics, \\
141 980 Dubna, Russia} 
\bigskip

\begin{abstract}
One of explanations
of the black hole entropy implies that 
gravity is entirely 
induced by quantum effects. By using arguments based on 
the AdS/CFT correspondence we
give induced gravity interpretation of the gravity in a brane
world in higher dimensional anti-de Sitter (AdS) space.
The underlying 
quantum theory is $SU(N)$ theory 
where $N$ is related to the CFT central charge. 
The theory includes massless fields
which correspond to degrees of freedom of the boundary CFT.
In addition, on the brane 
there are massive degrees of freedom
with masses proportional to $l^{-1}$ where $l$ is the
radius of AdS. 
At the conformal boundary of AdS
they are infinitely heavy and completely decouple.
It is the massive fields which can explain the black hole entropy. 
We support our interpretation by 
a microscopic model of a 2D brane world in 
$AdS_3$.  
\end{abstract}

\bigskip

\bigskip

\baselineskip=.6cm

\noindent
\newpage

The title of this talk\footnote{This talk was given 
at the International
Conference "Quantization, Gauge Theory and Strings"
dedicated to the memory of Professor E.Fradkin,
Moscow, June 5-10, 2000.} 
includes three different ideas. The first 
idea is that black holes possess thermodynamical properties 
which hint at microscopical degrees of freedom
responsible for the black hole entropy. The second one is 
Sakharov's idea \cite{Sakharov} that the Einstein action 
can be entirely induced by vacuum polarization effects in quantum
field theory on curved backgrounds. The rest part 
of the title is related to
new alternative to compactification \cite{ADD},\cite{RS} 
which treats
our world as a brane embedded in a space of higher 
dimension.  If curvature of this space is negative
the gravity localized on the brane is similar to 
the Einstein gravity \cite{RS}.  This alternative
may help to solve the hierarchy problem and it predicts a new 
physics in accelerator experiments.  
The purpose of this talk is to present, by joining these three ideas, 
arguments in favor of induced gravity interpretation
of gravity in brane worlds \cite{RS} and discuss features
of the corresponding microscopic theory\footnote{Recently 
the induced gravity approach was also discussed in \cite{DGP} for 
branes in extra Minkowski space. Here we consider higher 
dimensional space with negative curvature.}.  


According to black hole thermodynamics black holes possess an entropy 
$S^{BH}$
which is related to the surface area of the horizon ${\cal A}$. 
The precise relation is given by the Bekenstein-Hawking formula
$S^{BH}={\cal A}/(4G)$ where $G$ is the Newton constant.
This quantity appears in purely classical theory and it is, perhaps,
one of the strongest arguments that the Einstein gravity is
not fundamental but is an effective form of 
some underlying quantum theory 
which has microscopical degrees of freedom capable to explain
$S^{BH}$. At the present moment the string/D-brane theory is the best 
candidate to this role and there is a number of computations
that prove it correctly counts the black hole entropy (see 
\cite{Peet} for a review).  
Unfortunately direct computations in the string theory are rather
involved. That is why the above results concern
extremal and near extremal black holes where one can use
additional arguments and reduce the
problem to statistical mechanics of some simple dual systems.
Although these results are very important they are not quite 
satisfactory. 
First, the computations
are not universal and, second, they say little 
about true degrees of freedom in the physical space-time of the
black hole.

There are other 
explanations of $S^{BH}$. 
One of them \cite{Jacobson},\cite{FFZ} is based on Sakharov's idea that 
the Einstein action can be entirely induced by quantum effects.
The Bekenstein-Hawking entropy in induced gravity can be related
to loss of information about states of a physical
vacuum inside the black hole horizon (see \cite{F} for recent
discussion and references). This explanation is 
universal in a sense it is equally applicable
to Schwarzschild, Kerr or Kerr-Newman black holes 
and does not depend on details of the models \cite{FF}.  
The degrees of freedom which are responsible for $S^{BH}$ are
those quantum fields (constituents) which induce the Einstein
action. Their location 
is a narrow strip near the black hole horizon.

It should be emphasized that induced gravity models 
\cite{FFZ}--\cite{FF} are not necessarily an alternative to the string 
theory because the both pictures have common basic features.
In the string theory the classical gravity and 
matter fields appear from a tree-level closed string diagram
which, on the other hand, can be interpreted as a one-loop
diagram of an open string \cite{GSW}. Thus, 
there is a possibility that gravity and  
black hole entropy in string theory are purely 
quantum loop effects
\cite{HMS}.
  
To proceed let us consider a simple induced gravity model 
of \cite{FFZ}. It consists of free massive scalar and spinor 
fields. Suppose that masses $m_i$ of all constituents are determined
by a single scale $M$, i.e., that $m_i=\alpha_i M$
where $\alpha_i$ are dimensionless
coefficients of the order of unity.
The induced Newton constant $G$ can be computed exactly \cite{FFZ},
$G^{-1}=\gamma M^2$, where
\begin{equation}\label{2}
\gamma={1 \over 6\pi}\left[\sum_s(1-6\xi_s)\alpha_s^2\ln \alpha_s
+2\sum_d \alpha_d^2\ln \alpha_d\right]~~~
\end{equation}
and indices $s$ and $d$ are referred
to scalars or spinors, respectively. The constants $\xi_s$
determine non-minimal couplings of scalar constituents which are
required to make $G$ free from ultra-violet divergences.
Without special tuning the coefficient $\gamma$ is of the
order of $\alpha_i$ and to have the correct value for $G$ one
has to assume that $M$ is of the order of the Planck mass $M_{Pl}$.
 
Suppose we replace each field in the considered model by a
$SU(N)$ multiplet in adjoint representation. 
It is clear 
that in the new model $G^{-1}=\gamma (N^2-1) M^2$
and at large $N$ one has to take $M\sim M_{Pl}/N$.
This simple observation shows that there may be two scales in
the induced gravity,
one determined
by masses of fields and the other by their number. 
If $N$ is large, $M$ can be sufficiently
small and this may help to avoid the hierarchy problem. 

Consider now the brane world model suggested by Randall and Sundrum
\cite{RS}, more precisely, the scenario with single brane. 
We take the bulk cosmological constant $\Lambda=-6/l^2$ 
and assume that the brane divides the 
$AdS$ space into two identical parts.
The correspponding action is\footnote{We work in the Euclidean
theory.}
\begin{equation}\label{3}
I=-{1 \over 16\pi G_5}\int\sqrt{g}d^5x 
\left(^{(5)}R+12/l^2\right)+ 
{1 \over 4\pi G_5}\int\sqrt{h}d^4x (K+4\pi G_5\mu)~~~, 
\end{equation} 
Here $g$ and $h$ are
determinants of the bulk and the brane metrics, respectively,
$\mu$ is the brane tension. The extrinsic curvature 
$K$ appears in (\ref{3})
due to the jump on the brane
of normal derivatives of the bulk metric.
The role of the gravitational equations in the brane world
is played by the Israel equations
\begin{equation}\label{4}
{1 \over 4\pi G_5} (K_{\mu\nu}-h_{\mu\nu}K)
=\mu h_{\mu\nu}~~~.
\end{equation}
The solution of equations in the bulk 
can be written in the 
Fefferman-Graham form (see, for instance, \cite{Graham}
and references therein)
\begin{equation}\label{5}
ds^2_5={l^2 dr^2 \over r^2}+h_{\mu\nu}(r,x)dx^\mu dx^\nu~~~,
\end{equation}
where $r>0$ and $h_{\mu\nu}(r,x)$ satisfies certain constrains.
$h_{\mu\nu}(r,x)$ is singular at $r=0$ which is 
a conformal boundary of (\ref{5}).
The constraints on $h_{\mu\nu}$ can be resolved at small $r$
\cite{Graham}
\begin{equation}\label{6}
h_{\mu\nu}(r,x)={l^2 \over r^2}\left(g_{\mu\nu}(x)+
g^{(2)}_{\mu\nu}(x) r^2+
g^{(4)}_{\mu\nu}(x) r^4+
\bar{g}^{(4)}_{\mu\nu}(x) r^4\ln r/l+...\right)~~~.
\end{equation}
Here $g_{\mu\nu}$ is the metric of the conformal boundary,
$g^{(2)}_{\mu\nu}$ is linear in the curvature of $g_{\mu\nu}$,
$\bar{g}^{(4)}_{\mu\nu}$ is quadratic in the curvature
and traceless. Definition of 
$g^{(4)}_{\mu\nu}$ in terms of $g_{\mu\nu}$
admits an ambiguity and in general the two quantities are 
independent. Other terms in (\ref{6}) are determined
in  terms of $g_{\mu\nu}$ and
$g^{(4)}_{\mu\nu}$.

Suppose now that $r$ is the coordinate in the extra dimension
and the brane is located at a small fixed $r$. 
Hence, $x^\mu$ are
the coordinates on the brane. It follows
from (\ref{6}) that 
\begin{equation}\label{7}
{2 \over l}(K_{\mu\nu}-h_{\mu\nu}K)={6 \over l^2} h_{\mu\nu}
-(R_{\mu\nu}-{\frac 12}Rh_{\mu\nu})+
2r^2\ln {r \over l} ~\bar{g}^{(4)}_{\mu\nu}+U_{\mu\nu}~~~.
\end{equation}
Here $h_{\mu\nu}(x)$ is the metric on the brane which coincides with
$h_{\mu\nu}(r,x)$ at fixed $r$, $R$ and $R_{\mu\nu}$ are 
4D curvatures determined by $h_{\mu\nu}$. The tensor 
$U_{\mu\nu}$ is polynomial in curvatures of $h_{\mu\nu}$ and
depends on $g^{(4)}_{\mu\nu}$. 
By using (\ref{7}) and (\ref{4}) we conclude that 
gravity on the brane is an Einstein gravity, plus 
square curvature terms, plus a correction $U_{\mu\nu}$.
The four dimensional Newton constant is $G_4=G_5/l$.
To eliminate a cosmological constant in the brane world
one has to  put $\mu=3/(4\pi l G_5)$. 

The quantity $U_{\mu\nu}$ in (\ref{7}) is important 
because it results in 
a modification of the Newton law on the brane at small distances.  
There is an elegant interpretation \cite{Gubser} of this correction 
based on AdS/CFT correspondence \cite{M}. In a generalized formulation, 
this  
correspondence relates a classical gravitational
action on AdS
and the effective action of a quantum conformal field theory on the 
boundary of AdS. It enables one to relate $U_{\mu\nu}$
to the renormalized stress energy tensor
$T_{\mu\nu}$ of the CFT.
More precisely, if $U_{\mu\nu}$ is expressed in terms of
$g_{\mu\nu}$  and $r$
the quantity $8\pi G_4 T_{\mu\nu}(g)$
is the limiting value of $l^2/r^2 U_{\mu\nu}(r,g)$ 
when $r$ goes to zero.

This fact is crucial for our further interpretation:
if  gravity on the brane is really induced by quantum
effects then an underlying theory reduces to a conformal
boundary theory at vanishing $r$. 

We now have to understand how may this theory look like
at finite values of $r$ and what is the origin of the Einstein
term in equation (\ref{7}). 
We have two options:
a) there is a one-to-one correspondence of physical
degrees of freedom in
the theory at finite $r$ to degrees
of freedom of the boundary CFT and b) the underlying theory  
on the brane has extra degrees of freedom. 

The first option is possible when  theory at finite $r$ 
is treated as a regularized form of the CFT. 
In this case the parameter $r$ is related to the ultraviolet
cutoff of the CFT (the so called IR/UV duality) and
the local terms in (\ref{7}) are generated by 
divergences of the effective action. In the canonical
AdS/CFT they are eliminated by "renormalization"
of action (\ref{3}).

There is, however, a number of objections against this
point of view and the central argument is that
a regularized theory always
has unphysical properties. For instance,
in the dimensional regularization one formally works 
with complex dimensionalities, in the Pauli-Villars scheme
some fields have wrong statistics, a lattice regularization
breaks the Lorentz invariance.
There are also technical objections. 
Presumably, the dual CFT is ${\cal N}=4$ 
supersymmetric Yang Mills (SYM). This theory  cannot result in  
the divergent cosmological constant and explain the 
origin of $6 
l^{-2}h_{\mu\nu}$ in (\ref{7}) \footnote{To save the situation one may 
argue that at finite $r$ some degrees of freedom are massive and 
part of supersymmetries is broken.  
It cannot 
happen however
because at $r\rightarrow 0$ these massive fields would become 
infinitely heavy (see below)
and the resulting boundary theory would have less 
degrees of freedom than the CFT.}.  There may be also difficulties in 
explaining the Einstein tensor in (\ref{7}): when this theory 
is considered on a curved manifold it does not result in 
divergences of the Newton constant, at least at one loop.

We, thus, come to another option: the theory at finite $r$
is an ordinary QFT which includes extra physical degrees of freedom
as compared to the boundary CFT. The extra fields are
massive and induce the same local terms which appear in the
4D equations from Fefferman-Graham asymptotic.
The structure of the local terms in the 4D 
Lagrangian can be easily understood from (\ref{3}) and (\ref{6})
\begin{equation}\label{8}
{1 \over G_5} 
\sum_{n=0}l^{2n+1}(c R^{n+1}+bR^n\nabla^2 R+...)\sim
N^2\sum_{n=0}l^{2(n-1)}(c R^{n+1}+bR^n\nabla^2 R+...)
\end{equation} 
where $b$, $c$ are dimensionless coefficients. 
The quantity
$N^2= \pi l^3/(2G_5)$ appears in the AdS/CFT as
a coefficient by the conformal anomaly in the 
boundary CFT (a central charge) and is
related to the gauge $SU(N)$ group. The form of 
gravitational couplings
in (\ref{8}) indicates that
masses of the fields are of the order of $l^{-1}$
while their number is proportional to $N^2$.
These properties are typical to induced gravity
models \cite{FFZ}--\cite{FF}
where all fields are taken as adjoint $SU(N)$ multiplets.

To summarize, 
the underlying QFT looks as a 
$SU(N)$ gauge theory with both massive and
massless degrees of freedom. When the curvature radius on the brane
is much larger than $l$ the massive fields induce the
Einstein gravity while the massless fields result in 
a correction in a form of non-local
stress energy tensor $U_{\mu\nu}$, see (\ref{7}).
When one goes to the theory at zero $r$ the fates of 
massless and massive fields are different. Massless fields 
simply reduce to the boundary CFT. 
As for the massive fields, their masses on the boundary 
have to be rescaled in accordance with the
definition of the boundary metric $g_{\mu\nu}=r^2l^{-2} h_{\mu\nu}$.
This leads at zero $r$ to infinitely large
masses $lr^{-1}m\sim r^{-1}$. 
Hence, the massive fields do not propagate on the boundary and
all the theory reduces to the CFT.

It is clear that finding an induced gravity model
for a 4D brane world explicitly is a complicated problem.
So let us illustrate how this induced gravity scenario
is realized in  a simplified model of a two-dimensional
brane world in $AdS_3$. In three dimensions the bulk action 
has the same form as (\ref{3}) with 
cosmological constant $\Lambda= -l^{-2}$ and
Newton constant $G_3$. 
Consider one of solutions of bulk equations which can be found
by following the analysis of \cite{SS}
\begin{equation}\label{9}
ds^2={l^2 \over r^2}dr^2+{1 \over r^2}e^{-2\sigma}
\left((1+r^2a)^2d\tau^2+(1-r^2b)^2dx^2 \right)~~~,
\end{equation}
\begin{equation}\label{10}
a=\frac 14 l^2e^{2\sigma} (\sigma')^2~~,~~
b=\frac 14 l^2e^{2\sigma} (2\sigma''+(\sigma')^2)~~.
\end{equation}
Here $\sigma=\sigma(x)$, $\sigma'=d\sigma/dx$
and we suppose that now $r$ is dimensionless. Solution (\ref{10})
corresponds to the boundary metric $e^{-2\sigma}(d\tau^2+dx^2)$.
For simplicity we suppose that 
the brane is located at $r=1$ and its
metric is
\begin{equation}\label{11}
ds_b^2=e^{-2\sigma}
\left((1+a)^2d\tau^2+(1-b)^2dx^2 \right)~~.
\end{equation}
(Metrics on branes 
located at arbitrary $r$ are obtained from (\ref{11})
by trivial redefinition of $\sigma$.)
The Israel equations (\ref{4}) in two dimensions are reduced to
the single independent equation
\begin{equation}\label{12}
K_{\tau\tau}=-4\pi\mu G_3 e^{-2\sigma}(1-b)^2~~.
\end{equation}
Together with (\ref{9}),(\ref{10}) it yields $a=-b=-(z-2)/(z+2)$,
where $z\equiv 8\pi\mu lG_3$ is a parameter (definition
(\ref{10}) requires $z<2$). As a result, the "dilaton"
field $\sigma$ is described by the Liouville equation
\begin{equation}\label{13}
e^{2\sigma}\sigma''= {4 \over l^2}{z-2 \over z+2}
\end{equation}
and we can conclude that the brane-world geometry 
(\ref{11}) is locally 
$AdS_2$ with the curvature radius $l/(2-z)$. 
The Liouville gravity has a number of interesting properties,
one of which is the existence of solutions which can be
interpreted as $AdS_2$ black holes. The entropy of these
black holes can be explained in the corresponding
induced gravity model \cite{FFGK}. 
It is for this
reason, to have the Liouville gravity on the brane, we have chosen
the bulk solution in form (\ref{9}). Another alternative, a 
2D de Sitter brane, was discussed in \cite{HMS}.

Let us construct now an effective gravitational action 
on the brane, i.e., a functional $\Gamma[h]$
of the brane metric such that
\begin{equation}\label{15}
{2 \over \sqrt{h}}{\delta \Gamma[h] \over \delta h_{\mu\nu}}=
-{1 \over 4\pi G_3}(K^{\mu\nu}-h^{\mu\nu}K)~~~.
\end{equation}
It is more easy to "integrate" this equation if one brings 
brane metric (\ref{11}) to a conformal form
\begin{equation}\label{16}
ds_b^2=e^{-2\varphi}(d\tau^2+dy^2)~~~,
\end{equation}
\begin{equation}\label{17}
e^{-2\varphi}=e^{-2\sigma}(1+a)^2~~,~~
\left({dy \over dx}\right)^2=
\left({1-b \over 1+a}\right)^2~~
\end{equation}
and find dependence on $\varphi$ from equation
\begin{equation}\label{18}
{\delta \Gamma[\varphi] \over \delta \varphi}=
-{1 \over 4\pi G_3}e^{-2\varphi}K~~~.
\end{equation}
To integrate (\ref{18}) is still a difficult task
which can be accomplished 
if the "dilaton coupling" is small,
$e^{2\varphi}\ll 1$. The latter means that the brane is close to
the boundary. 
From (\ref{17}) one gets
\begin{equation}\label{19}
K=-{2 \over l}\left[1+{l^2 \over 2}\varphi'' e^{2\varphi}
+{l^4 \over 8}(\varphi')^2(2\varphi''+(\varphi')^2)
e^{4\varphi}+...\right]~~,
\end{equation}
where $\varphi'=d\varphi/dy$. A solution to (\ref{18})
is a series in $e^{2\varphi}$
\begin{equation}\label{20}
\Gamma[\varphi]=-{l \over 4\pi G_3}
\int d\tau dy e^{-2\varphi}
\left[{1 \over l^2}+
\frac 12 (\varphi')^2e^{2\varphi}
+{l^2 \over 24}
(\varphi')^4e^{4\varphi}+...\right]~~.
\end{equation}
As follows from (\ref{20}) 
the brane action, expressed in terms of the metric
$h_{\mu\nu}$ has a very non-local form. At small
couplings only the first two terms 
in (\ref{20}) are important and
$\Gamma$ coincides with action of the Liouville theory
with central charge $c=3l/G_3$. 
This fact is not surprising: reducing the coupling $e^{2\varphi}$
is equivalent to moving the brane closer to the boundary where
one recovers the boundary CFT, the
Liouville theory in the given case.

If $c$ is large the discussed Liouville gravity 
can be entirely induced  by a two-dimensional
ultraviolet finite
quantum field theory on a curved background with metric (\ref{16}),
see \cite{FFGK}. The theory includes $N$ 
free massless
scalar fields, where $N=3l/G_3$ 
and a number of massive  scalar and spinor fields which
induce the 2D cosmological term. 
It is not difficult to see how this model has to be
modified to induce the non-Liouville terms in
(\ref{20}). The simplest way is to introduce
"non-minimal" couplings of massive scalar constituents 
to the dilaton, $V_s(\varphi)\phi_s^2$,
and find out $V_s(\varphi)$ perturbatively. 
In the leading 
order 
\begin{equation}\label{21}
V_s=e^{4\varphi}\left[
\alpha_s (\varphi')^4+\beta_s(\varphi'')^2+O(e^{2\varphi})
\right]~~~,
\end{equation}
where $\alpha_s$, $\beta_s$ are dimensionless constants,
$\sum_s\alpha_s=\sum_s\beta_s=0$.
By a special choice of $\beta_s$ one can cancel
$R^2$-terms in the effective quantum action which are absent in 
(\ref{20}). $\alpha_s$ can always be chosen to reproduce the 
coefficient by $(\varphi')^4$ in (\ref{20}).

In this way one can compute $\Gamma$ and $V_s$ perturabatively
order by order and find out a non-local microscopic quantum theory 
which describes gravity in the 2D brane world. This theory
has the same features as we discussed above for the four-dimensional
brane. First of all, though the
central charge of the boundary Liouville theory is determined only
by the number of the massless degrees of freedom, the
presence of the massive fields is important for 
the same reasons as in four dimensions
(to cancel divergences and induce a cosmological constant). 
There are other common features. For instance, 
in 2D
the numbers of scalars and spinors, $N_s$ and $N_d$, 
are related to $N$ by the constraint $N_s+N-2N_d=0$ needed 
to cancel the divergences.
Thus, the  most natural way to go to large $N$ is to suppose that
$N_s\sim N_d\sim N$. 
A simple model with this property
is where all fields are in the fundamental representation of 
$SU(N)$. 
If  masses of the constituents are fixed
by a scale $M$  the 2D induced cosmological
constant $\lambda$ behaves as $NM^2$. 
On the other hand, in the brane world
model $\lambda\sim 1/(lG_3)$ (see (\ref{20})) 
and $N\sim l/G_3$, which means that
$M\sim l^{-1}$. Hence, as in four dimensions,
masses of fields should be proportional to the inverse radius of
$AdS_3$.

In the end let us return to the problem of the black hole entropy.
In a 2D brane world in the small dilaton coupling limit
this entropy can be interpreted as the
entropy of entanglement of massless constituents \cite{FFGK}.
This is, however, a feature of two dimensions.
The degrees of freedom responsible for
$S^{BH}$ in four dimensions
are massive constituents \cite{FFZ}. 
Massless constituents corresponding to the boundary CFT
are analogous to a quantum gas around a black hole
and their entropy should be identified with a quantum correction
to $S^{BH}$.  
This differs our interpretaion from the point of view 
of \cite{HMS}.

\bigskip

\noindent{\bf Acknowledgements.} The author is grateful to 
J. Buchbinder, N. Kaloper and G. Kunstatter for helpful
discussions.
This work is supported in part by the 
RFBR grant N 99-02-18146 and NATO Collaborative Linkage Grant 
CLG.976417.


\end{document}